\documentclass[aps,prc,a4paper,twocolumn,showpacs,showkeys,floatfix, superscriptaddress]{revtex4-1}

\usepackage[dvips]{epsfig}
\usepackage{color}
\usepackage{amsmath}
\usepackage{multirow}

\newcommand{\figwidth}{\columnwidth}
\newcommand{\dtn}{\textnormal{NiCl$_2\cdot$4SC(NH$_2$)$_2$}}

\begin{document}

\title{Antiferromagnetic resonance in a spin-gap magnet with strong single-ion anisotropy}

\author{V.N.Glazkov}
\email{glazkov@kapitza.ras.ru}
\affiliation{P.Kapitza Institute for Physical Problems RAS, 119334 Moscow, Russia}
\date{\today}

\begin{abstract}
Quasi-one-dimensional magnet \dtn{}, usually abbreviated as DTN, does not order at zero field down to $T=0$: due to the strong single-ion anisotropy of the ``easy plane'' type acting on  $S=1$ Ni$^{2+}$ ions, the  $S_z=0$ ground state is separated from  $S_z=\pm 1$ excitations by an energy gap. Once the magnetic field is applied along the main anisotropy axis, the gap closes at  $B_{c1}=2.18$~T and the field-induced antiferromagnetic order arises. The low-energy excitations spectrum of this field-induced ordered state includes two branches of excitations, one of them have to be a gapless Goldstone mode. Recent studies of excitations spectrum in a field-induced ordered state of DTN (T.Soldatov \emph{et.al}, Phys.Rev.B \textbf{101}, 104410 (2020)) have revealed that Goldstone mode became gapped as magnetic field deviates from the main symmetry axis. This paper proposes simple description of antiferromagnetic resonance modes of quasi-one-dimensional quantum $S=1$ magnet with strong single-ion anisotropy. The approach used is based on a combination of the strong coupling model for the anisotropic spin chain with the conventional mean-field model of antiferromagnetic resonance. The resulting model fits to the known experimental results without additional tuning parameters.
\end{abstract}


\maketitle

Metal-organic compound  DTN  (\dtn) is an example of the collective paramagnet with gapped excitations spectrum (spin-gap magnet). Magnetic ions Ni$^{2+}$ ($S=1$) form chains running along the four-fold axis of the tetragonal crystal \cite{dtn-struct,dtn-neutrons}. Contrary to Haldane magnets, magnetic ions in DTN are subjects of the  strong single-ion anisotropy, which separates  $S_z=0$ single-ion state from the excited doublet $S_z=\pm1$. Exchange coupling turns out to be weak as compared to the  single ion anisotropy, it leaves ground state nonmagnetic and makes $S_z=\pm 1$ excitations delocalized. Applied magnetic field reduces energy of one of the excited states and closes the gap at a certain critical field, leading at the same moment to the formation of the field-induced long-range order due to the weak inter-chain coupling \cite{paduan-old,paduan2004}.

Field-induced ordering of the spin-gap magnets was actively discussed in the literature in relation with Bose-Einstein condensation of magnons \cite{giamarchi,zapf-review}. One of expected properties of the field-induced ordered state is the existence of a gapless Goldstone mode due to the conservation of axial symmetry above the critical field, but the low symmetry of the real magnets can violate this prediction. Tetragonal symmetry of DTN crystals makes this compound one of the most suitable candidates for the search of the Goldstone mode in the field-induced ordered state.

Dynamics of low-energy excitations in DTN was studied by means of electron spin resonance (ESR) spectroscopy in Refs.~\cite{zvyagin,zvyagin-smirnov-glazkov,soldatov}. Recent study by Soldatov et. al. \cite{soldatov} has revealed that at small tilt (up to 5$^\circ$) of the applied field from the four-fold symmetry axis the Goldstone mode becomes gapped. ESR frequencies in the field-induced ordered phase of DTN were calculated using various theoretical approaches \cite{zvyagin-smirnov-glazkov,utesov}, however, these models do not yield compact expressions neither for the magnetic resonance eigenfrequencies nor for the dependencies of the ESR spectra parameters on exchange coupling and single-ion anisotropy constants. This work provides interpretation of the ESR eigenfrequencies in the field-induced ordered phase of the $S=1$ quantum magnet DTN, which allows to follow characteristic dependence of ESR eigenfrequencies on the parameters of microscopic model and describes quantitatively results obtained for the weakly tilted  magnetic field.

Spin Hamiltonian for DTN can be written as:

\begin{equation}\label{eqn:ham}
    {\cal H}_{\textrm{chain}}= \sum_i \left(D {  S}_{z,i}^2+ J{ {\vec S}}_i{ {\vec S}}_{i+1}-g\mu_B \vec{B} {\vec{S}}_i \right)
\end{equation}

\noindent for DTN $D=8.9$~K, $J=2.2$~K, strongest inter-chain exchange integral is approximately ten-fold smaller (0.18~K) and is omitted in (\ref{eqn:ham}), the $g$-factor value for the field applied along the tetragonal axis equals $g=2.26$ \cite{zvyagin}.

\begin{figure}
\epsfig{clip=, width=\figwidth,file=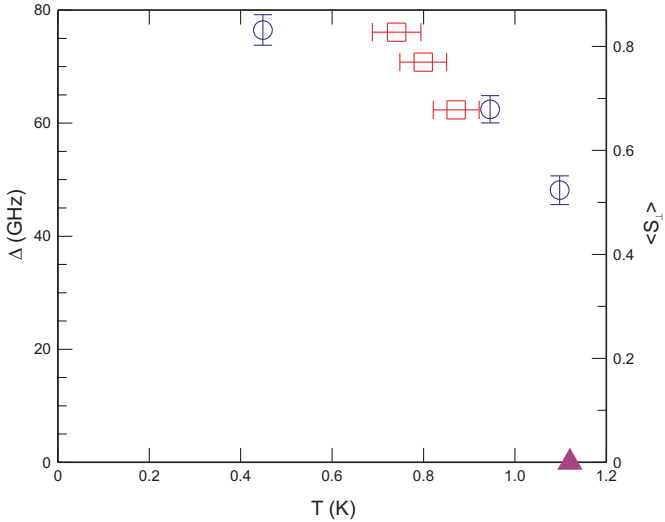}
\caption{Temperature dependence of the mean value of the ordered transverse spin component at the field 8~T, $\langle S_\perp \rangle$ values are calculated from experimental data using Eqn.~\ref{eqn:Delta}. Left Y-axis shows gap values $\Delta$ from Ref.~\cite{zvyagin-smirnov-glazkov}, right Y-axis shows the mean value of the ordered transverse spin component. Open symbols are $\Delta(T)$ data: squares corresponds to the temperature scans at the fixed frequency, circles corresponds to the $f(B)$ measurements at the fixed temperature. Filled triangle on X-axis is the temperature of phase transition at the field of 8~T according to Ref.~\cite{dtn-neutrons}. }
\label{fig:order}
\end{figure}

\begin{figure}
\epsfig{clip=, width=\figwidth,file=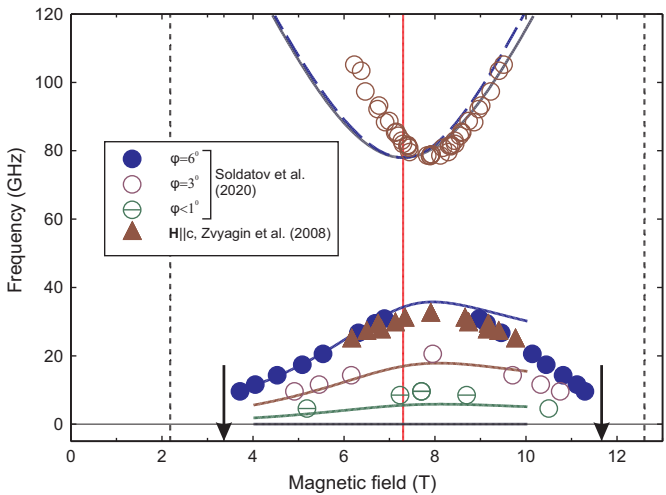}
\caption{Comparison of the model results (Eqn.~(\ref{eqn:f1-general})) with the experimental results of Ref.~\cite{soldatov} (circles) and Ref.~ \cite{zvyagin-smirnov-glazkov} (triangles). Vertical solid line marks  $B_0$ value, vertical dashed lines mark experimentally found values of the critical fields $B_{c1}$ and $B_{c2}$, arrows mark critical field values computed from Eqns.~(\ref{eqn:Bc1}),(\ref{eqn:Bc2}). Curves are model results for the Hamiltonian parameters given in the text. Curves for the lower ESR branch corresponds to the tilt angles (bottom to top) $0^\circ$, $1^\circ$, $3^\circ$ and $6^\circ$, correspondingly. Curves for the upper branch are computed for the tilt angles $0^\circ$ (solid curve) and $6^\circ$ (dashed curve).}
\label{fig:theory}
\end{figure}

In the strong-coupling limit $D\gg J$ dispersion of the spin excitations can be found within conventional perturbations theory for $\vec{B}||Z$ as long as the applied field is below the critical value. Zero-field excitations spectrum for the $S=1$ case is found up to third order over $J/D$ in Ref.~\cite{papa} (similar result is also known for the arbitrary spin \cite{exp}), magnetic field can be taken into account straightforwardly since $S_z$ remains a good quantum number. One-magnon excitations spectrum above the saturation field can be found exactly (Ref.~\cite{zvyagin} includes these calculations for the case of DTN accounting for the inter-chain couplings as well).

Minimal energy of these excitations at $(ka)=\pi$ turns to zero at critical fields \cite{dietrich,zvyagin}:
 \begin{eqnarray}
    g\mu_B B_{c1}&=&D-2J+\frac{J^2}{D}+\frac{J^3}{2 D^2}\label{eqn:Bc1}\\
    g\mu_B B_{c2}&=&D+4J\label{eqn:Bc2}
 \end{eqnarray}

Critical fields $B_{c1}$ and $B_{c2}$ are close within the strong coupling limit $D\gg J$. At $B_{c1}<B<B_{c2}$ single-ion Hamiltonian can be projected on the two lower states $S_z=0$ and $S_z=1$. This two-level system can be formally described by a pseudo-spin $T=1/2$ by relabeling single-ion states: $T_z=-1/2 \Leftrightarrow S_z=0$ and $T_z=1/2\Leftrightarrow S_z=1$ \cite{dietrich}. Spin operators have to be replaced as follows:
 \begin{eqnarray}
  {S}_z&=& {T}_z+1/2\\
  {S}^{\pm}&=&\sqrt{2} {T}^{\pm} \label{eqn:operators}
 \end{eqnarray}

After this substitution   Hamiltonian (\ref{eqn:ham}) transforms at $\vec{B}||Z$ into:

 \begin{eqnarray}
    {\cal H}&=&2J \sum_i \left({  T}_{x,i}{  T}_{x,i+1}+{  T}_{y,i}{  T}_{y,i+1}+\frac{1}{2}{  T}_{z,i}{  T}_{z,i+1}\right)+\nonumber\\
    &+&\left(J+D-g\mu_B B\right)\sum_i {  {{T}}}_{z,i}+N\frac{2J+D-g\mu_B B}{2}
    \label{eqn:ham-pseudo}
 \end{eqnarray}

\noindent this transformation has linear accuracy over $J/D$.

Thus, the problem of the $S=1$ spin chain with strong single-ion anisotropy is mapped onto the equivalent model of uniform $S=1/2$ XXZ-chain with strong XY-anisotropy and the effective field $B_{eff}=B-(J+D)/(g\mu_B)$. Effective field turns to zero at the applied field value $B_0=(J+D)/(g\mu_B)$, which is within first order of perturbations theory equal to the half-sum of the critical fields. One can also note that frequencies of the ESR-active transitions at $k=0$ at $B<B_{c1}$ within this approximation are given by ${h f=(D+2J+\frac{J^2}{D}-\frac{J^3}{2 D}\pm g\mu_B B)}$, extrapolation of the lower frequency turns to zero at ${B_0^{\textrm{(PM)}}=\left(D+2J+\frac{J^2}{D}-\frac{J^3}{2 D}\right)/(g\mu_B)}$, the later value is close to  $B_0$ and $B_{c1}$ but does not coincides with either of them. Similarly, extrapolation of the high-field ESR frequency (${k=0}$) at $B>B_{c2}$, ${h f=(g\mu_B B-D)}$, turns to zero at the field value ${B_0^{\textrm{(HF)}}=D/(g \mu_B)}$ \cite{zvyagin} different from both $B_0$ and $B_{c2}$ .

In terms of pseudo-spins inter-chain coupling also transforms into interaction with strong XY-anisotropy. Thus, at  $B=B_0$ and $T=0$ quasi-one dimensional $S=1$ magnet can be mapped on an equivalent model of the ordered 3D $S=1/2$ antiferromagnet with strong ``easy plane'' anisotropy in zero effective field.

Once the applied field deviates from $B_0$ (but remains parallel to $Z$), the equivalent model becomes that of an ``easy plane'' antiferromagnet in a field ${B_{eff}=B-B_0}$, applied along the anisotropy axis $Z$. Critical fields  $B_{c1}$ and $B_{c2}$, symmetrically located to the left and to the right from $B_0$, are the saturation fields for the equivalent model. ESR eigenfrequencies (antiferromagnetic resonance eigenfrequencies) for the ``easy plane'' antiferromagnet with strong anisotropy can be found within the conventional sublattices model \cite{nagomiya,goorevich}: for ${\vec B}_eff||Z$ one of the eigenfrequencies is zero, and the second eigenfrequency depends on the effective field as $f=\sqrt{(\gamma B_{eff})^2+\Delta^2}$, here $\gamma=g\mu_B/h$ is a gyromagnetic ratio. The first mode corresponds to the expected Goldstone mode, and the second mode corresponds to the gapped spectral branch observed in \cite{zvyagin-smirnov-glazkov,soldatov} (the gap $\Delta$ equals 78~GHz at $T=0.45$K).

Within the sublattices model the gap $\Delta$ can be derived from the $XY$-anisotropy of the equivalent model (\ref{eqn:ham-pseudo}) and magnitude of the antiferromagnetic order parameter (sublattice magnetization) \cite{dietrich,nagomiya,goorevich}:

 \begin{equation}
 \Delta=\gamma \sqrt{2 H_A H_E},
\end{equation}

\noindent here $H_E=4 J \mu/(g^2\mu_B^2)$ is the characteristic exchange field, $H_A=J\mu/(g^2\mu_B^2)$ is the characteristic anisotropy field for the equivalent model and $\mu$ is the  average sublattice magnetization for the equivalent model, the inter-chain couplings are neglected here. Hence $\Delta= 2\sqrt{2} (J/h) \langle t_\perp \rangle$, here $\langle t_\perp \rangle$ is the mean \emph{pseudo-spin} projection on the XY-plane. Transforming this result back to the real spins (\ref{eqn:operators}) one obtains expression for the gap of the upper ESR branch in terms of transverse component of the \emph{real} spin

\begin{equation}\label{eqn:Delta}
    \Delta=\frac{2 J}{h} \langle S_\perp \rangle
\end{equation}

\noindent Thus, one can conclude that magnitude of the gap $\Delta$ for the upper ESR branch in the ordered phase is mostly determined by inter-chain exchange integral $J$, while the position of the minimum of this branch (field $B_0$) is mostly determined by the single-ion anisotropy constant $D$.

Gap $\Delta$ was measured as a function of temperature in Ref.~\cite{zvyagin-smirnov-glazkov}. These data allows to recover temperature dependence of the order parameter $\langle S_\perp \rangle$ at the field of 8~T (Fig.~\ref{fig:order}). The result is qualitatively sound: obtained values of the order parameter $\langle S_\perp \rangle<1$.

Now we will allow applied magnetic field to tilt slightly from the $Z$ direction, this tilt towards $X$ axis can be parameterized by an angle $\Theta$. At small tilt angles field-induced antiferromagnetic ordering persists, while the critical field are slightly rescaled \cite{kalita}. Linearly over $\Theta$, Zeeman contribution to the pseudo-spin Hamiltonian (\ref{eqn:ham-pseudo}) becomes:

\begin{equation}\label{eqn:Zeeman-tilt}
    {\cal H}_Z=-g\mu_B B_{eff} \sum_i  {  T}_{z,i}-\sqrt{2} g \mu_B B \Theta \sum_i  {  T}_{x,i}
\end{equation}

\noindent i.e., additional contribution arises which is proportional to the \emph{full applied field} and is applied within the ``easy plane''. This contribution is assumed to be small: ${g\mu_B B\Theta \simeq D \Theta\ll J\ll D}$.

When the applied field is equal to $B_0$, the $Z$-component of the effective field is zero and our equivalent pseudo-spin model is that of an ``easy plane'' antiferromagnet with the field applied within the ``easy plane''. Eigenfrequencies of such an antiferromagnet  \cite{nagomiya,goorevich} are $f_1=\gamma B_\textrm{in-plane}$ and $f_2=\Delta$. Thus, one obtain quantitative result for the ESR frequency of the lower branch at the field $B_0\approx (B_{c1}+B_{c2})/2$:

\begin{equation}\label{eqn:f1-B0}
    f_1(\Theta)=\sqrt{2}\gamma B_0 \Theta
\end{equation}

\noindent Note, that contrary to the interpretation of the gap for the upper ESR branch, the later result is independent from the magnitude of the order parameter in the field-induced ordered phase.

If the amplitude of the applied field deviates from $B_0$, resulting effective field will be applied at some angle within the $XZ$-plane: $\vec{B}_{eff}=(\sqrt{2} B \Theta;0; (B-B_0))$  and one can use known expression for the antiferromagnetic resonance eigenfrequencies  \cite{nagomiya}
\begin{equation}\label{eqn:f-general}
    \frac{(\gamma B_{eff,x})^2}{f^2}+\frac{(\gamma B_{eff,z})^2}{f^2-\Delta^2}=1
\end{equation}
\noindent hence, for the resonance frequencies

\begin{widetext}
\begin{equation}
    f_{1,2}^2=\frac{1}{2} \left[\Delta^2+2(\gamma B \Theta)^2+\gamma^2 (B-B_0)^2\pm\sqrt{\left(\Delta^2+2(\gamma B \Theta)^2+\gamma^2 (B-B_0)^2\right)^2-8\Delta^2 (\gamma B\Theta)^2}\right]\label{eqn:f1-general}
\end{equation}
\end{widetext}

The model above, in fact, relies on a classical mean-field treatment of the two-sublattices antiferromagnet. Within the classical mean-field approach magnetization process is  linear in field up to the saturation field. Due to one-dimensionality of DTN magnetization changes non-linearly approaching the critical fields \cite{paduan2004,zapf-review,zvyagin}, and, moreover, this nonlinearity is not fully symmetric on approaching $B_{c1}$ from above and $B_{c2}$ from below. This asymmetry of the real magnet properties  compared to the model predictions is due to the limited accuracy of the linear over $J/D$ approximation used for mapping of the real spin problem (\ref{eqn:ham}) on a simpler pseudo-spin problem (\ref{eqn:ham-pseudo}). Thus, the proposed model is applicable to DTN only in certain vicinity of $B_0$, the range of model applicability can be roughly estimated as a field range with linear $M(B)$ dependence, which extends \cite{zvyagin} from 4 to 10~T. To go beyond these limits would require both to take into account higher orders on $J/D$ when transforming to pseudo-spin representation, and to take into account one-dimensionality of DTN spin subsystem. However, one can be certain that the frequency of the lower ESR branch will turn to zero at the critical fields $B_{c1}$ and $B_{c2}$, which are the saturation fields of the equivalent model.

Comparison of the model curves with experimental data from Ref.~\cite{soldatov} is shown at Fig.~\ref{fig:theory}. Model curves are computed without additional fitting parameters using the microscopic Hamiltonian (\ref{eqn:ham}) constants $D=8.9$~K, $J=2.2$~K, $g=2.26$ ($\gamma=31.6$~GHz/T) \cite{zvyagin}, which yields $B_0=7.3$~T (half-sum of experimentally measured values  of $B_{c1}$ and $B_{c2}$ equals 7.4~T), and experimentally measured value $\Delta=78$~GHz \cite{zvyagin-smirnov-glazkov,soldatov}. Model curves were computed for the tilt angle $\Theta$ equal to $0^\circ$, $1^\circ$, $3^\circ$, $6^\circ$, as was used in the experiment \cite{soldatov} (note that the practical accuracy of tilt angle determination is around $\simeq 1...2^\circ$). One can see, that the model curves agree well with the experimental data.

Summing it up, this paper proposes simple and illustrative model to describe magnetic resonance eigenfrequencies in the field-induced antiferromagnetically ordered phase of the quasi-one dimensional magnet with strong ``easy-plane'' anisotropy  \dtn{}. The proposed model provides simple relations between ESR eigenfrequencies and microscopic Hamiltonian parameters, it is also applicable for the case of magnetic field slightly tilted from the main anisotropy axis.

The work was supported by the Russian Science Foundation Grant No.~17-12-01505 and by the Program of RAS Presidium ``Actual problems of low-temperature physics''. Author thanks Prof.~A.I.Smirnov and Dr.~T.A.Soldatov (Kapitza Institute) for numerous fruitful discussions.

\end{document}